\documentclass[doublecol,comment]{epl2}
\usepackage{graphicx,epsfig,color}

\title{Reply to Comment by M. Rohloff, T. Lapp and J. Vollmer}

\author{Michael Wilkinson}

\institute{
{Department of Mathematics and Statistics,
The Open University, Walton Hall, Milton Keynes, MK7 6AA, England}
}

\begin{document}

\maketitle

\section{Introduction}
\label{sec: 1}

Reference \cite{Wil14} discusses a theory describing experiments on a binary mixture of
liquids close to a miscibility transition, with critical temperature $T_{\rm c}$.
In these experiments periods of turbidity
are observed as the system is driven away from the critical point by a slow
change of the temperature $T$. The turbidity is caused by precipitation of droplets which
drift towards the interface between the two phases due to their difference in density.
Reference \cite{Wil14} presents a theory for
the time period $\Delta t$ between these precipitation events, as a function of the rate of
change of the temperature of the system, described by a parameter $\xi$ with
dimensions of inverse time. The theory assumes that the initial growth of droplets
is by Ostwald ripening (as discussed in {\cite{Lif+61}), and that this crosses over to
collisional growth driven by gravitational
settling. The theory predicts that
\begin{equation}
\label{eq: 1}
\Delta t \sim K\, \xi^{-3/7}
\ .
\end{equation}
The coefficient $K$ is a function of the reduced temperature, $\theta=|T-T_{\rm c}|/T_{\rm c}$:
\begin{equation}
\label{eq: 2}
K(\theta)=\alpha \left(D\Lambda \kappa^3\right)^{-1/7}
\end{equation}
where $\alpha$ is a dimensionless constant, $D$ is the interdiffusion coefficient,
$\Lambda $ is a Kelvin length for the phase boundary and the settling speed of a droplet
of radius $a$ is $u=\kappa a^2$. The quantities $\Lambda$ and $\kappa$ are defined
in terms of tabulated material material parameters of the mixture, such as the interfacial
tension $\gamma$ and the density difference between the phases, $\Delta \rho$. Because
$\Delta \rho$, $\gamma$ and $D$ all approach zero as a power-law in $\theta$ at the critical point,
it is predicted that $K\sim \theta^{-\eta}$. Using standard critical exponents gives $\eta\approx 0.5$
\cite{Wil14}.

In \cite{Wil14} it is asserted that (\ref{eq: 1}) is a good description of the experimental
data, for the water/isobutoxyethanol system.
This statement is based upon a plot in the Ph.D. thesis of T. Lapp \cite{Lap11}, figure 3.26,
which is a scatter plot of $\Delta t\xi^{3/7}$ against temperature. The experimental points
collapse onto a curve which shows good agreement with equations (\ref{eq: 1}) and (\ref{eq: 2}), over a wide range of values of $\xi $ and $\theta$, with the dimensionless number
$\alpha$ close to unity.

The comment by Rohloff, Lapp, and Vollmer, however, is phrased
so as to suggest that their experimental results do not support the theory.
In the following I argue that the data which they present are fully in accord
with the theory as described in \cite{Wil14}. Furthermore, the basis of the claim
that there is a disagreement results from an attempt to extend the theory in
\cite{Wil14} by arguing that it can be used to determine a bound on the oscillation
period. I give two arguments indicating that this proposed extension of the theory
is unsound.

\section{Comparison with experimental data}
\label{sec: 2}

Figure 1 of the Comment is a version of figure 3.26 in \cite{Lap11}.
I am grateful to the authors of the Comment
for making this available in a widely disseminated journal, because it lends experimental
support to \cite{Wil14}.

In particular, the experimental data in figure 1 of the Comment indicate that
$\Delta t\xi^{3/7}$ is a function of temperature, as the theory predicts,
with both $\xi$ and $\theta$ varying over approximately two decades.
Furthermore, the scatter of data points for different values of $\xi$
appears to be random, indicating that there is no residual systematic dependence upon $\xi$.
This is very strong support for the scaling relation, equation (\ref{eq: 1}).

Concerning the temperature dependence of $K=\Delta t\xi^{3/7}$, the experimental
data points in figure 1 differ from the theoretical prediction (dotted green curve) by
what appears to be a constant offset. This constant offset on a logarithmic scale
corresponds to the unknown dimensionless constant $\alpha$ in (\ref{eq: 2}).
Moreover, the fact that the data points follow a line of constant slope in this double-logarithmic
plot is a indication that the prefactor $K$ has a power-law dependence on the reduced
temperature, $K\sim \theta^{-\eta}$, with $\eta \approx 0.5$. These observations
show that $K(\theta)$ is well-approximated by equation (\ref{eq: 2}).

The presentation of the experimental data described in the Comment
is a non-trivial test of the theory, and
the data in figure 1 of the Comment are entirely consistent with
the prediction in \cite{Wil14}.

One aspect of the presentation of figure 1 of the Comment should to be remarked upon.
The theoretical curve is not quite the same as that proposed in \cite{Wil14}, namely
equation (\ref{eq: 1}) above: an additional factor of $2.4396\times (4^3/3^4)^{1/7}\approx 2.4$
has been included in the theoretical expression used in the comment, namely equation (3)
of that work. This exaggerates the apparent \lq disagreement' with the experimental
data, as presented in figure 1 of the Comment.

\section{Proposed bound on the period}
\label{sec: 3}

The model developed in \cite{Wil14} includes a \lq collision
efficiency' $\epsilon$, which was assumed to be of order unity.
The authors of the Comment argue that $\epsilon <1$, and that this
inequality leads to a lower bound on the period $\Delta t$.
They argue that the resulting bound is violated by their experimental data.

There are two flaws in this argument. First, droplet growth in a system
which combines gravitational settling and Ostwald ripening is an extremely
complicated problem, involving a population of droplets with a distribution
of radii $P(a,t)$. This distribution is determined by a supersaturation field satisfying an advection
diffusion equation with moving boundaries (due to gravitational settling of the droplets).
The simplified model described in equation (11) of \cite{Wil14}, equivalently equation (1)
of the Comment, is a caricature of this complex system in terms of the evolution of
the radius of a single droplet, $a(t)$. While this has proven
sufficient to surmise scaling properties (as was done in \cite{Wil14}),
it is not adequate to establish quantitative bounds on the oscillation period (as proposed
in the Comment). Adapting equation (1) of the Comment
to yield a precise bound on the period is an unsound procedure.

Secondly, the argument is based upon the notion that a collision
efficiency $\epsilon$ cannot exceed unity. This inequality cannot be guaranteed for the
collision efficiency which is defined in \cite{Wil14}.
The value of $\epsilon$ is defined by considering a droplet of radius $a$ falling
through a gas of much smaller droplets with volume fraction $\Phi$ at a
relative speed $u$. The rate of increase of the volume $v_{\rm d}$ of the
droplet is defined to be
\begin{equation}
\label{eq: 3}
\dot v_{\rm d}=\epsilon \pi a^2 u \Phi
\ .
\end{equation}
In the case where the large droplet grows by coalescence upon contact,
the collision efficiency cannot exceed unity (and it may be substantially
smaller, as is thought to be the case for water droplets in clouds \cite{Mas57}).
In \cite{Wil14}, however, droplets grow by Ostwald ripening, which is
effected by diffusive transfer of material between droplets which need not be in contact.
A large droplet creates a reduction of the supersaturation field in its neighbourhood,
which causes material to evaporate from smaller droplets and condense on the larger droplet \cite{Lif+61}.
This mechanism allows a falling droplet to collect material from a cylindrical region which has a greater
area than its geometrical cross-section, implying that it is possible, in principle,
for the collision efficiency defined in (\ref{eq: 3}) to exceed unity.

While the collision efficiencies for water droplets in air have been extensively
investigated, those of small droplets in the the water/isobutoxyethanol system
do not appear to have been studied. Moreover the physics is fundamentally different
(as discussed in \cite{Wil14})
because the experiments are performed close to the critical point (down to $\theta\approx 10^{-3})$. There is, however, an argument supporting the assumption
that the collision efficiency $\epsilon$ is of order unity. If $\epsilon$ were
very small or very large, it is unlikely that it would be a constant. Instead, $\epsilon$ would be
expected to have a systematic dependence upon parameters of the model, which would
imply that the predicted scaling relations (\ref{eq: 1}) and (\ref{eq: 2}) would be
expected to fail. The success of these predictions is indirect evidence that $\epsilon \approx 1$.

\section{Summary}
\label{sec: 4}

The experimental data presented in the Comment are fully consistent with
the theory in \cite{Wil14}. Figure 1 of the Comment confirms the $\Delta t\propto \xi^{-3/7}$ scaling, and that the temperature dependence of the prefactor is in accord with the theory. The dimensionless
prefactor $\alpha$ would be close to unity if a comparison were made with the theory
in the form presented in \cite{Wil14}.

The claim that the theory in \cite{Wil14} does not agree with the experimental results is
a consequence of an inappropriate attempt to extend it. In particular, the model
used in \cite{Wil14} is not sufficiently accurate to support quantitative bounds
on the period, and the physics of the growth process does not guarantee that
$\epsilon=1$ is an upper bound on the \lq collision efficiency'.

\end{document}